\documentclass[prb,showpacs,twocolumn,floatfix]{revtex4}
\usepackage{graphicx}
\usepackage{dcolumn}
\usepackage{bm}
\begin{document}
\newcommand{\vk}{\textbf{k}}
\newcommand{\vq}{\textbf{q}}
\newcommand{\vp}{\textbf{p}}
\newcommand{\vx}{\textbf{x}}
\title{ Theoretical study of electronic Raman scattering of Borocarbide superconductors}
\author{Hyun C. Lee}
\email{hyunlee@phys1.skku.ac.kr}
\author{Han-Yong Choi}
\affiliation{BK21 Physics Research Division and Institute of Basic Science,
 Department of Physics,\\
Sung Kyun Kwan University, Suwon, 440-746
Korea}
\date{\today}
\begin{abstract}
The electronic Raman scattering of Borocarbide superconductors is studied based on the weak coupling  theory
with $s+g$-wave gap symmetry.
The low energy behaviors and the relative peak positions can be naturally understood,
while the explanation of the detailed shape of the $B_{1g}$  peak seems to require  a strong inelastic interaction not present  in the weak coupling
theory.  
\end{abstract}
\maketitle
\section{Introduction}
Electronic Raman scattering is a very useful probe in determining
the symmetry of the order parameter of superconductors (SC).\cite{cooper}
They play especially important roles in the study of unconventional superconductors owing to
 the strong dependence of Raman responses on the symmetry of superconducting order parameters.
Along with the discovery of  high temperature superconductors, other unconventional
superconductors have been actively investigated, for example,
Borocarbide superconductors (BCSC) \cite{first}
$\textrm{(Y,Lu)} \textrm{Ni}_2 \textrm{B}_2 \textrm{C}$, and 
Ruthenate superconductors\cite{ichida}  $\textrm{Sr}_2 \textrm{Ru}\textrm{O}_4$.
The symmetry of the order parameter of high temperature  SC and Ruthenate SC is believed 
to be $d$-wave\cite{annett} and $p$-wave\cite{ichida}, respectively.
However, the symmetry of the order parameter of BCSC has not been determined unambiguously at present.
Nevertheless, there are growing experimental evidences for the strongly anisotropic nature of 
 order paramter of BCSC.
The band structure calculations\cite{pickett,jilee} and $T_c$ versus $\gamma$ plot \cite{carter}
suggest  strong electron-phonon interaction as a
possible origin of BCSC.   In general, the superconductivity mediated by electron-phonon interaction has an isotropic 
$s$-wave superconducting gap (SG), and the nodes of  gap on Fermi surface (FS) do not exist. 

On the other hand, the following experiments indicate the existence of  the nodes of SG:
(1)  the power law  temperature dependence ($T^3$) of nuclear relaxation rate \cite{nmr}
$1/T_1$ suggests the existence of line nodes of the superconducting gap on Fermi surface.\cite{sigrist}
Moreover,  Hebel-Slichter peak  is not observed \cite{nmr}, which is also consistent with
the existence of nodes.
(2) The magnetic field ($H$) dependence of residual density of states in the vortex state\cite{nohara1,nohara2} 
was observed to be  $\sqrt{H}$. 
According to the well-known result by Volovik\cite{volovik}, the $\sqrt{H}$ indicates the existence of
lines of gap nodes on FS.
(3)  V. Metlushko \textit{et al.} reported an observation of the fourfold symmetric upper critical field.\cite{hc2} 
Evidently, such a behavior is not compatible with the isotropic $s$-wave gap, and  has been interpreted
in terms of \textit{three} dimensional $d$-wave superconductivity.\cite{maki}
(4) The \textit{direct} measurement of SG by photoemission spectroscopy\cite{yokoya} 
also demonstrates  the existence of nodes of SG.
(5) The recent experiment by E. Boaknin \textit{et al.} on the thermal conductivity of $\textrm{LuNi}_2 \textrm{B}_2\textrm{C}$
shows that the magnetic field dependence of thermal conductivity at low temperature is very similar to that
of a heavy fermion SC with gap which has \textit{lines of nodes}.\cite{boaknin}

The above experimental facts (1)-(5) clearly corroborate a gapless superconductivity, and the simplest of which is 
$d$-wave superconductivity. 
However, the following experiments for the BCSC with Ni substituted by some impurity element as Pt pose  serious challenges for the 
$d$-wave scenario:
(6) The magnetic field $H$ dependence of residual density of states in the vortex state\cite{nohara1,nohara2}
changes from $\sqrt{H}$ to $H$ upon the impurity substitution.
Caroli \textit{et al.}\cite{caroli} showed that  the $H$ dependence is linear for the isotropic $s$-wave SC.
(7) The detailed plot of $\ln [C(T)/\gamma T_c]$ versus $T_c/T$ also demonstrates the change of behavior from $T^3$ to 
the activated one.\cite{nohara3}
(8) The photoemission experiment shows the opening of gap over the whole FS for impurity substituted compound
by comparing the shapes of the spectra of pure and dirty BCSC.

The experiments for the dirty BCSC  (6)-(8) imply the gap  symmetry cannot be a simple d-wave because 
the introduction of impurities does not change the gap anisotropy  owing to the d-wave symmetry\cite{norman}, and
only  overall  amplitudes decrease with the impurity concentrations. 
In connection with the experiments for the diry BCSC we note that the
 superconducting temperature of conventional $s$-wave SC is insensitive to the introduction of non-magnetic impurities,
which is the fact known as Anderson theorem\cite{anderson}, while the unconventional SC 
 depend on them sensitively.\cite{norman,larkin}

Finally there is no experimental evidence of broken tetragonal symmetry, thus we are forced to rule out pairing
interactions which violate tetragonal symmetry.

Synthesizing all of the above observations, (1)-(5), (6)-(8), and the unbrokent tetragonal symmetry,
we are led to a proposal that the gap symmetry of BCSC should be a mixed one
which respects the tetragonal symmetry. 
Obviously, the  simplest possible candidate is the $s+g$-wave gap symmetry.  
The gap function of the form $\Delta(\phi)=|\Delta_0 \cos(2\phi)|$ is also a viable choice, but the nodes of the gap of this kind are accidental and
are not enforced by symmetry.  From the symmetry viewpoint, $s+g$-wave gap is more natural.

 Yang \textit{et al.} studied \cite{yang} BCSC by means  of electronic Raman scattering.
They observed $2 \Delta$-like peaks in $ A_{1g}, B_{1g}$, and $B_{2g}$ geometry.
 The peak position of $B_{1g}$ geometry is found to be larger than the others.
Also, the  peaks of the $A_{1g}, B_{2g}$ symmetry are much stronger than the peak of $B_{1g}$ symmetry.
There exists scattering strength below $2 \Delta$ gap, which increases with frequency 
\textit{linearly} in $B_{1g}, B_{2g}$ geometry. [See Fig. 5]

We intend to understand the qualitative features of the above Raman experiments based on
the simplest model with $s+g$-wave gap symmetry. We adopt the weak coupling theory with separable pairing interaction.
Then, even if the  physical origin of pairing interaction is unknown, we can deduce many physical consequences which 
mainly depend on the pairing symmetry not on the details of interactions.
Strictly speaking, the weak coupling theory is not expected to valid on a quantitative level, and indeed the band theory
calculation predicts the electron-phonon interaction to lie in the strong coupling regime.
Nevertheless, we believe that the qualititave features might be understood with the simple weak coupling model.

We find that the relative peak positions  and the nature of scattering strengh below $2 \Delta$ gap can be
understood naturally, while the weak and broad $B_{1g}$ peak features cannot be explained in  a simple weak coupling theory
with $s+g$-wave  gap symmetry.
Presumably, in $B_{1g}$ geometry,  other inelastic scattering mechanisms in particle-hole channel become important.  
The $B_{1g}$ phonon may provide one of such scattering mechanisms.

This paper is organized as follows:
In Sec. II,  we briefly summarize basic formalisms of electronic Raman scattering. In Sec. III,  the explict expressions of 
Raman intensity of $s+g$-wave SC are derived.
The physical results are presented in Sec. IV, and we conclude this paper with summary and discussion in Sec. V.
\section{Formalism}
The general formalisms for the electronic Raman scattering of superconductors are  well expounded in
the papers by Klein and Dieker\cite{klein}, Monien and Zawadowski \cite{monien}, and Devereaux and Einzel\cite{devereaux1}. 
Here we shall  briefly summarize the main results of the above papers necessary for our discussions.

The electronic Raman cross section is proportional to the  dynamical structure factor
$ S(\omega,\vq )$:
\begin{equation}
S(\omega,\vq )=[1+n_B (\omega )]\,\Big[-\frac{1}{\pi} \textrm{Im} \chi_{\tilde{\rho} \tilde{\rho}}^R(\omega,\vq )
\Big],
\end{equation}
where $n_B(\omega)$ is the Bose distribution function and the superscript $R$
of $\chi_{\tilde{\rho} \tilde{\rho}}^R$ denotes the retarded correlation function.
The effective density-density correlation function $ \chi_{\tilde{\rho} \tilde{\rho}}$ (in imaginary time) is defined by
\begin{equation}
\chi_{\tilde{\rho} \tilde{\rho}}(\tau-\tau^\prime,\vq)
=-\langle T_\tau \tilde{\rho}_\vq(\tau) \tilde{\rho}_{-\vq}(\tau^\prime) \rangle.
\end{equation}
$\tilde{\rho}_\vq$ is the effective density operator
\begin{equation}
\tilde{\rho}_\vq=\sum_{\vp,\sigma} \gamma_\vp \, c^\dag_{\vp+\vq/2,\sigma} \, c_{\vp-\vq/2,\sigma}.
\end{equation}
For  non-resonant electronic Raman scatterings,  the coefficient $\gamma_\vp$ is given by
\begin{equation}
\gamma_\vp=\sum_{\alpha \beta}\,e^I_\alpha \frac{\partial^2 \epsilon_\vp}{\partial p_\alpha \partial p_\beta} e^F_\beta,
\end{equation}
where $e^I, e^F$ is the polarization vector of the incoming and outgoing photons, respectively.
$\epsilon_\vp$ is the energy dispersion of material. $\alpha, \beta$ label the coordinates perpendicular to the photon momentum.

Light couples with the charge fluctuations in the material, and in this case the long-range Coulomb interaction should be
summed  in random phase approximation (RPA).  The RPA summation implements the screening effect. 
Let us define \textit{irreducible} correlation functions which do \textit{not} include the RPA type diagrams as subdiagrams.
\begin{eqnarray}
\pi_{\gamma \gamma}(\vq,\tau-\tau^\prime) &=&-\langle T_\tau \tilde{\rho}_\vq(\tau) 
\tilde{\rho}_{-\vq}(\tau^\prime) \rangle_{\textrm{irr}}, \\
\pi_{\gamma 0}(\vq,\tau-\tau^\prime) &=&-\langle T_\tau \tilde{\rho}_\vq(\tau) 
\rho_{-\vq}(\tau^\prime) \rangle_{\textrm{irr}}, \\
\pi_{0 \gamma}(\vq,\tau-\tau^\prime) &=&-\langle T_\tau \rho_\vq(\tau) 
\tilde{\rho}_{-\vq}(\tau^\prime) \rangle_{\textrm{irr}}, \\
\pi_{00}(\vq,\tau-\tau^\prime) &=&-\langle T_\tau \rho_\vq(\tau) 
\rho_{-\vq}(\tau^\prime) \rangle_{\textrm{irr}},
\end{eqnarray}
where $\rho_\vq=\sum_{\vp,\sigma} 1 \,c^\dag_{\vp+\vq/2,\sigma}\, c_{\vp-\vq/2,\sigma}$ is an ordinary density operator.
Then the correlation function $\chi_{\tilde{\rho} \tilde{\rho}}$ can be written as
\begin{equation}
\label{susceptibility1}
\chi_{\tilde{\rho} \tilde{\rho}}=\Big[ \pi_{\gamma \gamma}-\frac{\pi_{\gamma 0} \pi_{0 \gamma}}{\pi_{00}}\Big ]+
\frac{\pi_{\gamma 0} \pi_{0 \gamma}}{\pi^2_{00}}\,\chi_{\rho \rho},
\end{equation}
where $\chi_{\rho \rho}$ is  the density-density correlation function which is negligible in the low $\vq$ limit.
In the low $\vq$ limit, only the first term in the bracket of Eq. (\ref{susceptibility1}) needs to be considered.
\begin{equation}
\label{reducedsusc}
\lim_{\vq \to 0} \chi_{\tilde{\rho} \tilde{\rho}}(i\omega,\vq)
= \pi_{\gamma \gamma}(i\omega,\vq)-\frac{\pi_{\gamma 0}(i\omega,\vq) \pi_{0 \gamma}(i\omega,\vq)}{\pi_{00}(i\omega,\vq)}.
\end{equation}
The second term of the right hand side of Eq. (\ref{reducedsusc}) 
is usually referred to as the screening correction.

The vertex factor $\gamma_\vp$ for the lattice with tetragonal symmetry in  various scattering geometries are given by
\begin{eqnarray}
\label{vertexfactor}
\gamma_{A_{1g}}(\phi)&=&1+\gamma_A \cos 4 \phi, \nonumber \\
\gamma_{B_{1g}}(\phi)&=&\gamma_{B_1} \cos 2 \phi, \nonumber \\
\gamma_{B_{2g}}(\phi)&=&\gamma_{B_2} \sin 2 \phi.
\end{eqnarray}
There can be terms with higher harmonics such as 
$\cos(4 N \phi), \cos((4N-2) \phi), \sin((4N-2)\phi)$ with $N \ge 2$
in Eq. (\ref{vertexfactor}). We will ignore them, but in  the detailed fitting of  experimental
data,  the higher harmonics terms are necesary.\cite{devereaux1}
The factor of $1$ for the $A_{1g}$ geometry case of Eq. (\ref{vertexfactor}) is cancelled  by the screening correction,
therefore, effectively $\gamma_{A_{1g}}(\phi) \to \gamma_A \cos 4 \phi $.

The coefficients $\gamma_A, \gamma_{B_1}, \gamma_{B_2}$ are determined by the shape of  energy band.
If we take a simple band structure with  tetragonal symmetry [ $a$ is lattice constant]
\begin{eqnarray}
\epsilon_\vp&=&-2t [ \cos (p_x  a)  + \cos (p_y  a) ]  + 4 t^\prime \cos(p_x a) \cos(p_y a)  \nonumber \\
&-&2 t_z \cos(p_z a).
\end{eqnarray}
then  $\gamma_{B_{1g}} \propto t $ , $\gamma_{B_{2g}} \propto t^\prime$ and
$\gamma_A \propto  \textrm{const.} t  + \textrm{const.} t^\prime$.
In our work, we will simply put $\gamma_A=\gamma_{B_1}=\gamma_{B_2}=1$.  Thus, there are some reservations in comparing
the relative intensities of theoretical results with those of experimental ones.
\section{The $s+g$-wave superconductor}
The gap function of $(s+g)$-wave superconductor is
\begin{equation}
\label{thegap}
\Delta(\phi)=\Delta_s+\Delta_g \cos (4 \phi),\;\;  0< \Delta_s < \Delta_g,
\end{equation}
where $\phi$ is the azimuthal angle in spherical polar coordinate. The weak dependence on the radial momentum is 
neglected ($k=k_F$).  
Since the amplitude of $g$-wave part is larger than $s$-wave part the \textit{lines} of nodes 
exist on the FS. In the opposite case $\Delta_s > \Delta_g$ the gap function does not have nodes, and it behaves like
gapful anisotropic $s$-wave SC. 
The case $0< \Delta_s < \Delta_g$  and  $0< \Delta_g < \Delta_s$ 
is physically relevant for  pure BCSC and  dirty BCSC, respectively.
We will use the Nambu formalism for the computation of correlation functions in superconducting state.
The one-particle Green function in weak coupling approximation   is given by
\begin{equation}
\hat{G}(i\epsilon,\vp)=\frac{i \epsilon \tau_0 + \xi_\vp \tau_3 + \Delta_\vp \tau_1}{(i \epsilon)^2-
\xi_\vp^2-\Delta^2_{\vp}},
\end{equation}
where $\tau_i$'s are the Pauli matrices in Nambu space.
The wave function renormalization $Z(i\epsilon)$ is neglected, and the frequency dependence of interaction is encoded in the
cutoff   $\omega_c$   [See Eq. (\ref{pairing})] of pairing potential in  momentum space.
In Nambu notation, the correlation functions  $\pi_{\gamma \gamma^\prime}$ can be expressed as
[ $p=(i\epsilon,\vp), q=(i\omega,\vq)$]
\begin{equation}
\label{susceptibility}
\pi_{\Gamma \gamma^\prime}(q)=\frac{T}{N} \sum_{p} \textrm{Tr} \Big[
\hat{\Gamma}_{p,p+q} \hat{G}(p)\,\hat{\gamma^\prime}_{\vp+\vq} \hat{G}(p+q)\Big],
\end{equation}
where $N$ is the number of lattice sites.
$\hat{\Gamma}_{p,p+q}$ is the renormalized vertex function. 
We will evaluate the vertex function in ladder approximation.
The unrenormalized (bare) vertex function is
\begin{equation}
\hat{\Gamma}_{p,p+q}=\hat{\gamma}_\vp=\tau_3 \gamma_\vp.
\end{equation}
The gap equation of (spin-singlet) $s+g$-wave SC reads
\begin{equation}
\label{gapeq}
\Delta(\phi)=-\frac{1}{N}\,\sum_{\vp^\prime} D_{\vp \vp^\prime}\,\frac{\tanh (E_{\vp^\prime}/2T)}{2 E_{\vp^\prime}}\,
\Delta(\phi^\prime),
\end{equation}
where $D_{\vp \vp^\prime}$ is the pairing potential in momentum space, and 
$E_{\vp^\prime}=\sqrt{\xi_{\vp^\prime}^2+\Delta^2_{\vp^\prime}}$.

The gap equation Eq. (\ref{gapeq}) can be solved in a closed form for separable pairing interactions. 
Given the specific form of gap Eq. (\ref{thegap}), the separable pairing interaction which respects the tetragonal symmetry should be
chosen as 
\begin{eqnarray}
\label{pairing}
D_{ \vp \vp^\prime}&=&[ D_s + D_g \cos 4\phi   \cos 4 \phi^\prime \nonumber \\
&+&D_{sg} ( \cos 4\phi+ \cos 4 \phi^\prime )] \nonumber \\
&\times &\Theta(\omega_c-|\xi_{\vp}|) \Theta(\omega_c-|\xi_{\vp^\prime}|),
\end{eqnarray}
where $\Theta(x)$ is the  step function.
The numerical values of the coupling constants $D_s, D_{sg}, D_g$
are determined to be consistent with the values of $\Delta_s, \Delta_g$,  and the superconducting
transition temperature $T_c$.  For details, see Appendix C.
Since there is no dependence on the polar angle $\theta$ in  Eq. (\ref{susceptibility}), 
the momentum integration can be written as
\begin{equation}
\frac{1}{N} \sum_\vp=N_F \int d \xi_\vp  \frac{1}{2\pi} \int_0^{2\pi},
\end{equation}
where $N_F$ is the density of states per spin. Note that the spin sum is implicit in the trace over Nambu space.
For later references, let us define the dimensionless coupling constants.
\begin{equation}
\lambda_s=-N_F D_s,\;\lambda_g=-N_F D_g,\;\lambda_{sg}=-N_F D_{sg}.
\end{equation}
The technical analysis of Eq. (\ref{susceptibility}) crucially depends on whether the bare or the renormalized vertex is used.

\subsection{Analysis for the bare vertex}
Henceforth, we will consider only zero temperature case for simplicity. 
For the bare vertex, the integrals over frequency and $|\vp|$  of Eq. (\ref{susceptibility}) can be done explicitly.
Then the imaginary parts of $\pi^{R}_{\gamma \gamma^\prime}$ can be expressed as a single integral over azimuthal angle $\phi$.
\begin{equation}
\label{novertex}
-\textrm{Im} \pi^{R}_{\gamma \gamma^\prime}=
\frac{4 N_F}{\omega} \textrm{Re} \Big[ \int_0^{2\pi} \frac{d \phi}{2\pi} \frac{ \gamma(\phi) \gamma^\prime(\phi) 
\Delta^2(\phi)}{ \sqrt{\omega^2-4 \Delta^2(\phi)}}\Big].
\end{equation}
At finite temperature the factor $\tanh(\omega/4T)$ should be multiplied to the above result.

At this point we note that the screening correction is absent for $B_{1g}$ and $B_{2g}$ symmetry.
The screening correction can re-expressed as
\begin{equation}
\label{screen}
\pi_{\gamma 0}=N_F \int_0^{2 \pi}  \frac{d \phi}{2\pi}\,[\gamma(\phi) \cdot 1] \Delta^2(\phi) f(\cos 4 \phi),
\end{equation}
where $f$ is a certain complex function which can be expanded in a power series with respect to argument.
Clearly the integrals of Eq. (\ref{screen}) vanish for the vertex $\gamma(\phi)=\cos 2\phi$ or 
$\gamma(\phi)= \sin 2\phi$, since
they are orthogonal to $\cos 4 N \phi $ with $ N \ge 1$.
Thus, the screening corrections for $B_{1g}$ and $B_{2g}$ symmetry vanish.  
The $A_{1g}$ vertex  $\cos(4\phi)$  has a finite overlap with the rest of integrand,
and there is non-zero  screening correction for $A_{1g}$ symmetry.
The integrals Eq. (\ref{novertex}) cannot be done in  closed form, and the numerical integrations are required.
The results are presented in Sec. IV A.

\subsection{Analysis for the renormalized vertex}
\label{sub:vertex}
The vertex correction  will be evaluated within ladder approximation.
We will include the corrections coming from the pairing interaction only. In principle, other interactions may well contribute to the
vertex correction. However, we do not have enough experimental informations to identify the specific interactions responsible
for scatterings in particle-hole channel. 
Evidently this approximation should be improved, and in fact, from the comparison with the Raman experiment we expect the 
existence of strong  inelastic scatterings with $B_{1g}$  symmetry in particle-hole channel.
Here  we choose to work in the simplest approximation of including the pairing interaction only.

The summation of ladder diagrams is equivalent to solving an integral equation of the following type:
\begin{eqnarray}
\label{ladder}
& &\hat{\Gamma}_\gamma(\vp+\vq/2,\vp-\vq/2,i \omega)=
\hat{\gamma}( \vp+\vq/2,\vp-\vq/2) \nonumber \\
& & +\frac{T}{N}\,\sum_{i \epsilon^\prime,\vp^\prime}\,[(-1) D_{\vp \vp^\prime}]
 \tau_3 \,\hat{G}(\vp^\prime-\vq/2,i\epsilon^\prime)\,\nonumber \\
&  &\times  
\hat{\Gamma}_\gamma(\vp^\prime+\frac{\vq}{2},\vp^\prime-\frac{\vq}{2},i \omega)  \,
\hat{G}(\vp^\prime+\frac{\vq}{2}, i\epsilon^\prime+i\omega)\,\tau_3.
\end{eqnarray}
The retardation effect in weak coupling approximation  is reflected in the cutoff of pairing  interaction
in \textit{ momentum space}, and then the  frequency channel can be treated as instantaneous.
For such interaction, the vertex correction $\hat{\Gamma}$ becomes independent of incoming
frequency $i\epsilon$, which is an extremely crucial simplification.
In fact, for a fully self-consistent treatment, the renormalized Green function $\hat{G}$  with $(Z(i\epsilon) \neq 1)$ 
should be used,  which incoporate the
self-energy correction due to the the scatterings in particle-hole channel. 
However, if the underlying system is Fermi liquid, $Z(i \epsilon)$ correction is not expected to introduce qualitative changes.  

The momentum transfer $\vq$ can be neglected for SC with large penetraton depth.\cite{klein,monien,devereaux1}
In the limit $\vq=0$, it can be shown that \cite{klein} 
\begin{equation}
\hat{\Gamma}_\gamma(\vp,i \omega)=\Gamma_{2 \gamma}(\vp,i \omega)\tau_2+\Gamma_{3 \gamma}(\vp,i \omega)\tau_3.
\end{equation}
Then  Eq. (\ref{ladder}) simplifies to [$p^\prime=(i\epsilon^\prime,\vp^\prime)$]
\begin{eqnarray}
\label{vertexcorrection}
\Gamma_{2 \gamma}(\vp,i \omega)&=&\frac{T}{N} \sum_{\vp^\prime,i \epsilon^\prime}\,
\Big[ -\Gamma_{2 \gamma}(\vp^\prime,i \omega)\,A(p^\prime, i \omega) \nonumber \\
&+& 
\Gamma_{3 \gamma}(\vp^\prime,i \omega) B(p^\prime, i \omega) \Big] \times [-D_{\vp \vp^\prime}\,]\nonumber \\
\Gamma_{3 \gamma}(\vp,i \omega)&=&\gamma_\vp+\frac{T}{N} \sum_{\vp^\prime,i \epsilon^\prime}\,
\Big[ \Gamma_{3 \gamma}(\vp^\prime,i \omega)\,C(p^\prime, i \omega) \nonumber \\
&+& 
\Gamma_{2 \gamma}(\vp^\prime,i \omega) B(p^\prime, i \omega) \Big]\times[-D_{\vp \vp^\prime}\,],
\end{eqnarray}
where 
\begin{eqnarray}
\label{integralvertex}
A(p^\prime,i\omega)&=&
\frac{i\epsilon^\prime(i\epsilon^\prime+i\omega)-\Delta^2_{\vk^\prime}-\xi_{\vk^\prime}^2}{
((i\epsilon^\prime)^2-E_{\vk^\prime}^2)(
(i\epsilon^\prime+i\omega)^2-E_{\vk^\prime}^2)},\nonumber \\
C(p^\prime,i\omega)&=&
\frac{i\epsilon^\prime(i\epsilon^\prime+i\omega)-\Delta^2_{\vk^\prime}+\xi_{\vk^\prime}^2}{
((i\epsilon^\prime)^2-E_{\vk^\prime}^2)(
(i\epsilon^\prime+i\omega)^2-E_{\vk^\prime}^2)},\nonumber \\
B(p^\prime,i\omega)&=&
\frac{ i \Delta_{\vk^\prime}\,i\omega}{((i\epsilon^\prime)^2-E_{\vk^\prime}^2)(
(i\epsilon^\prime+i\omega)^2-E_{\vk^\prime}^2)}.
\end{eqnarray} 
The symmetry structure of pairing potential Eq. (\ref{pairing}) and the separable kernel form of 
Eq. (\ref{vertexcorrection}) imply that the vertices for $B_{1g}$ and $B_{2g}$ symmetry are not
renormalized, namely:
\begin{equation}
\Gamma_{2 \gamma}(\vp,i\omega)=0, \,  \Gamma_{3 \gamma}(\vp,i\omega)=\gamma_{\vp}, \;\;\textrm{for}\;\;
B_{1g}, B_{2g}.
\end{equation}
If the additional scatterings were present in particle-hole channel with $B_{1g}$ and/or $B_{2g}$  symmetry
the vertices for $B_{1g}$ and $B_{2g}$ symmetry would be renormalized. 

The  non-trivial correction influences only the $A_{1g}$ vertex  without the additional interactions mentioned above.
The vertex correction for the $A_{1g}$ can be calculated explicitly. The details of the calculations
can be found in Appendix B.
\section{Results}
\subsection{Results with the bare vertex}
\begin{figure}[hp!]
\scalebox{0.7}{\includegraphics{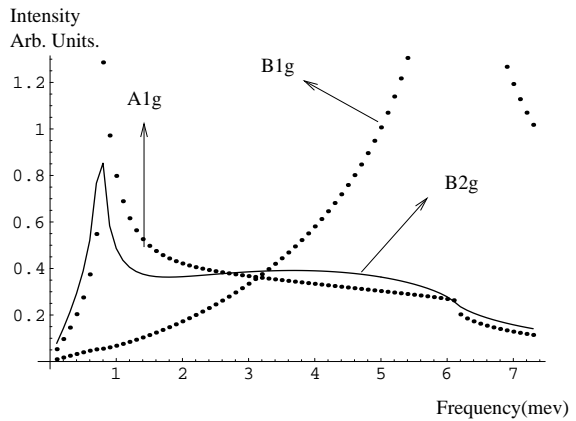}}
\caption{\label{novertexresult} Raman Intensities with the bare vertex. 
 $\Delta_g=1.72 \textrm{mev}$, $\Delta_s=1.34 \textrm{mev}$, and $T_c= 15.9 \textrm{K}$.}
\end{figure}

\begin{figure}[hp!]
\scalebox{0.7}{\includegraphics{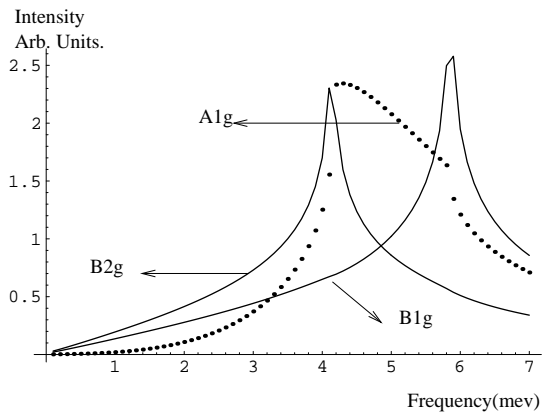}}
\caption{\label{experimentfit} Raman Intensities with the bare vertex.
$\Delta_g=2.5 \textrm{mev}$, $\Delta_s=0.43\textrm{mev}$, and $T_c=14.2 \textrm{K}$.}
\end{figure}

Even though  the analytic solutions in closed forms are not  available, 
the asymptotic behavior at low frequency and the peak positions of 
correlation functions can be understood without numerical integrations. 
[For a pure $d$-wave SC, the integral can be done exactly in terms of elliptic integrals.\cite{devereaux1}]
The detailed analytic calculations of low energy behaviors of  Raman susceptibilities $\pi_{\gamma \gamma}$ can be found
in Appendix A.
Numerical integrations are required for the solutions valid over the  entire frequency range.  
The numerical results are presented in 
Fig. \ref{novertexresult}  and Fig. \ref{experimentfit}. 
The result shown in Fig. \ref{novertexresult} is for the case $\Delta_g$ is slightly larger than $\Delta_s$, while
Fig.  \ref{experimentfit} shows the result for the case $\Delta_g$ is much larger than $\Delta_s$.

The $B_{1g}$ spectrum is characterized by a sharp peak at $\omega=2 (\Delta_s+\Delta_g)$,  and a linear frequency dependence at low frequency.
The low frequency behavior is described by [See  Eq. (\ref{lowB1case})]
$(1-\Delta_s/\Delta_g) \omega + \textrm{const.} \omega^3$.
Note that the slope of linear term is larger for smaller $\Delta_s/\Delta_g$.

The $A_{1g}$ and $B_{2g}$ spectra are characterized by  a peak at $\omega=2 (\Delta_g-\Delta_s)$, and a 
linear frequency dependence at low frequency.
Without the screening correction the $A_{1g}$ spectrum would have peaks at \textit{both} $\omega=
2 (\Delta_s+\Delta_g)$ \text{and} $\omega=2 (\Delta_g-\Delta_s)$. [See discussions below.]
The peak at $\omega=2 (\Delta_s+\Delta_g)$ of $A_{1g}$ spectrum is  cancelled by screening correction, however, 
a small hump feature remains at the frequency.
 $B_{2g}$ spectrum does not show noticeable structutre at  $\omega=2 (\Delta_s+\Delta_g)$.

The peak positions of the above spectra are determined by the following three conditions.
(A)  The angular integrals  Eq. (\ref{novertex}, \ref{screen}) can be re-expressed in terms of $z=\cos 4\phi$, 
then the Jacobian of the change of 
variable  $1/\sqrt{1-z^2}$ appears in the integrand. Then, 
large contributions  to the integrals stem from  the angular region  where $z=\cos 4\phi  \sim \pm 1$.
This feature is very similar to that of a van Hove singularity.
(B)  Large contributions to the integrals also come from the angular region where
$\omega^2-4 \Delta^2(\phi) \sim 0$. [ See Eq. (\ref{novertex}) ]
The frequencies which satisfy the conditions (A) and (B) are  $\omega=2 (\Delta_g \pm \Delta_s)$,
and the peaks are expected at those frequencies. 
(C) The  conditions (A) and (B) 
predict peaks at  \textit{both} frequencies $\omega=2 (\Delta_g +\Delta_s), 2 (\Delta_g -\Delta_s)$ 
irrespective of the scattering symmetry. 
The scatttering symmetry is encoded in the vertex factor $\gamma(\phi)$. 

For $B_{1g}$ symmetry,  $\gamma^2(\phi)=\cos^2(2\phi)=\frac{1+\cos 4\phi}{2}$.
The vertex  ( squared ) vanishes at $\cos 4\phi=-1$, and this factor suppresses a peak that would appear
at $\omega=2 (\Delta_g -\Delta_s)$. [ Recall $\Delta(\phi)=\Delta_s+\Delta_g \cos 4\phi$ ]
Therefore, for  $B_{1g}$ symmetry, only a single peak at $\omega=2 (\Delta_g+\Delta_s)$ is expected, which agrees with
the numerical result. 

The vertex factor of $B_{2g}$ is $\gamma^2(\phi)=\sin^2(2\phi)=\frac{1-\cos 4\phi}{2}$. Following the
same arguments as the  $B_{1g}$ case, we may expect a single peak at
 $\omega=2 (\Delta_g -\Delta_s)$, which also agrees with the numerical result.
 The screening corrections do not alter the above results because the corrections simply vanish for $B_{1g}, B_{2g}$ spectra
as discussed in Sec. III A.

For the  $A_{1g}$ case, we have to take the  screening correction into account.
Now notice that in the angular region  $\cos 4\phi \sim 1$, [ See Eq. (\ref{vertexfactor}) ]
\begin{equation}
\pi_{\gamma_{A_1} \gamma_{A_1}} \sim \pi_{\gamma_{A_1} 0} \sim \pi_{0 0}.
\end{equation}
Then, the contributions to the  Raman susceptibility, 
\begin{equation}
\chi_{\gamma_{A_1} \gamma_{A_1}} =
\pi_{\gamma_{A_1} \gamma_{A_1}}-( \pi_{\gamma_{ A_1}  0} )^2/\pi_{0 0},
\end{equation}
coming from the angular region  $\cos 4\phi \sim 1$
 becomes small due to the cancellation
between the first and the second term.
The above cancellations \textit{also} occur for the contributions from the angular region $\cos 4 \phi \sim-1$.
Then,  subleading contributions determine the peaks.  Numerical results show that the $A_{1g}$ peak appear 
at $\omega=2(\Delta_g-\Delta_s)$. 

\subsection{Results with the renormalized vertex}
\begin{figure}[hp!]
\scalebox{0.7}{\includegraphics{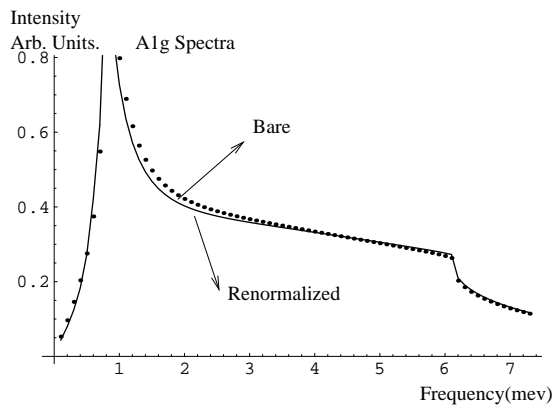}}
\caption{\label{comparefig} Raman intensity with/without vertex correction for $A_{1g}$ spectra.
$\Delta_s=1.34 \textrm{mev}$,  $\Delta_g= 1.72 \textrm{mev}$, and $T_c=15.9 \textrm{K}$.
$\lambda_s=0.13, \lambda_g=0.22, \lambda_{sg}=0.15.$  The solid line is the spectra with the vertex corretion, and the dotted line is  the 
bare vertex spectra.}
\end{figure}

\begin{figure}[hp!]
\scalebox{0.7}{\includegraphics{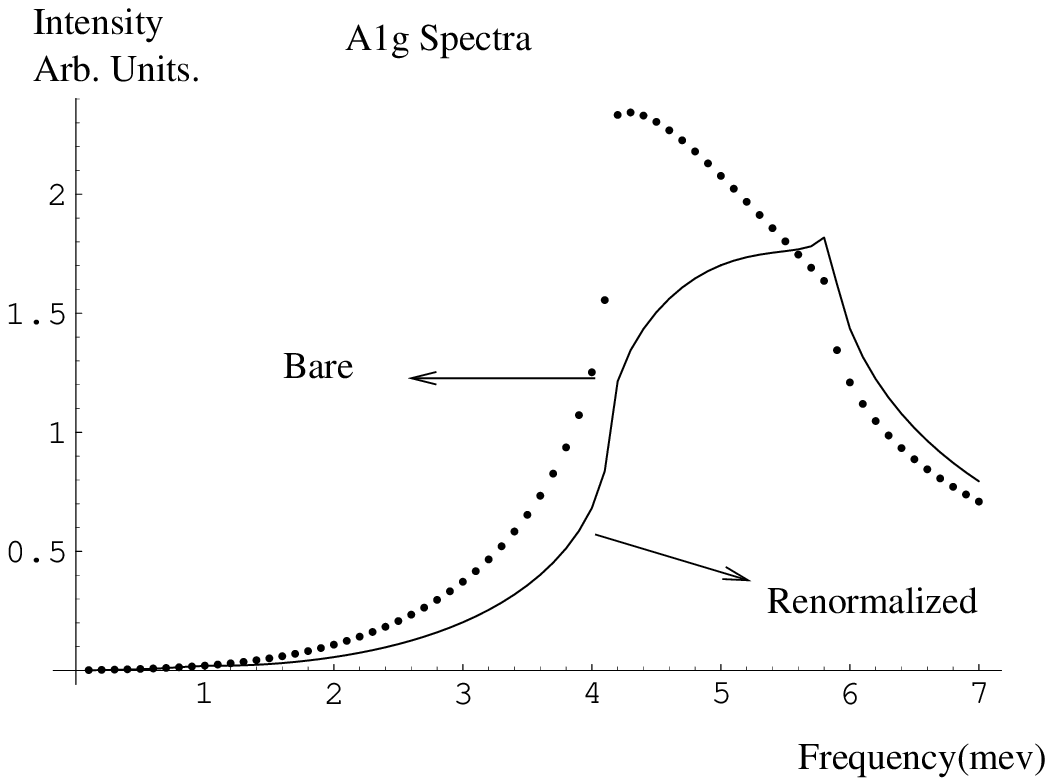}}
\caption{\label{vertexresult} Raman Intensity with vertex correction for $A_{1g}$ spectra.
$\Delta_s=0.43  \textrm{mev}$,  $\Delta_g= 2.5 \textrm{mev}$, and $T_c=14.2 \textrm{K}$.
$\lambda_s=0.05, \lambda_g=0.42, \lambda_{sg}=0.06$.
The solid line is the spectra with the vertex corretion, and the dotted line is  the 
bare vertex spectra.}
\end{figure}

As discussed in Sec. III B, the vertex correction affects only  $A_{1g}$ spectrum.
The vertex correction depends on the relative magnitude of $\Delta_s$ and $\Delta_g$. 
The results are presented in Fig. \ref{comparefig} and Fig.  \ref{vertexresult}.
It turns out that the vertex correction is more  important  for larger $\Delta_g/\Delta_s$. 
In case of Fig. \ref{comparefig}  $[\Delta_g/\Delta_s=1.28]$, the vertex correction is almost negligible over the entire frequency range.
On the other hand, in case of Fig. \ref{vertexresult} $[\Delta_g/\Delta_s=5.8]$,   the vertex correction suppresses the peak
at $\omega=2(\Delta_g-\Delta_s)$ which was present in the spectrum with the bare vertex.  At low frequency, the vertex correction is negligible
as in  the case of Fig. \ref{comparefig}, while at frequency higher than $2 ( \Delta_g+\Delta_s)$, it slightly enhances the spectrum.
More discussions will follow in the next section.
\section{Discussions and Summary}
\begin{figure}[hp!]
\scalebox{0.7}{\includegraphics{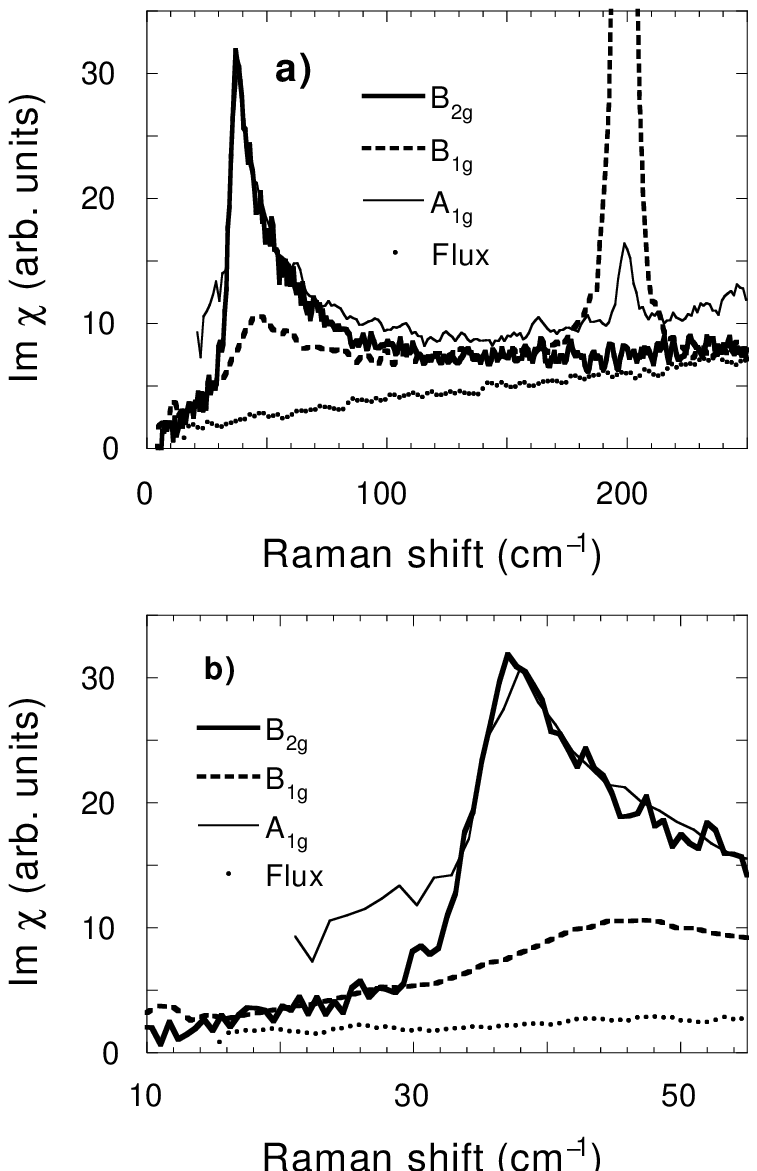}}
\caption{\label{yangfig} Raman spectra of $\textrm{YNi}_2 \textrm{B}_2 C$ with $T_c=15.3 \textrm{K}$
  from Yang  \textit{et al.}'s experiment \cite{yang}. The strong peak of (a) near 200 $\textrm{cm}^{-1}$ is 
due to $B_{1g}$ phonon. }
\end{figure}

Comparing the theoretical results, especially Fig. \ref{experimentfit} , with the experimental data  Fig. \ref{yangfig} (b), we find that
 the relative order of peak positions  and the low frequency behaviors coincide with those of experimental data.
In our model, the peak of $B_{1g}$ spectrum is located at $\omega=2 (\Delta_s+\Delta_g)$, while that of
$B_{2g}, A_{1g}$ is located at $\omega=2 (\Delta_g-\Delta_g)$. Then
 $\Delta_g$ and $\Delta_s$ can be extracted from the experimental data, and we obtain
$\Delta_g \sim 2.5 \textrm{mev}$ and $\Delta_s \sim 0.3 \textrm{mev}$. 
Thus, it turns out that the $s$-wave component of gap is much smaller than the $g$-wave component.
In fact, Fig. \ref{experimentfit} [$\Delta_g/\Delta_s=5.8$] compares more favorably with the experimental data  Fig. \ref{yangfig} 
than Fig. \ref{novertexresult} [$\Delta_g/\Delta_s=1.28$] .

We have to note that the linear frequency  dependence   $B_{1g}$ spectrum  at low frequency  observed in the Raman experiment
rules out the  pure $d$-wave scenario,
which predicts $\omega^3$ dependence. [See Appendix A] 
The low frequency behaviors are  robust againt the effect of vertex correction  and 
provide an unambiguous evidence  for the $s+g$-wave scenario.

But the theoretical results  pose some drawbacks. 
Most significantly, the shape of $B_{1g}$ peak looks very different from the experimental one.
Theoretical results predict a strong  \textit{sharp} peak at $\omega=2 (\Delta_g+\Delta_s)$. 
Experimentally, the $B_{1g}$ peak is very broad and weak compared to  $A_{1g}, B_{2g}$ peaks.

However, there exist some caveats in comparing the theoretical results directly with the experimental data.
Recall that we have included only the pairing interaction in particle-hole channel  in computing  the vertex correction.
Owing to the specific symmetry of pairing interaction, the $B_{1g}$ and $B_{2g}$ channels were immune from the vertex correction.
In general, we may  well include the interactions in  $B_{1g}$ and $B_{2g}$ symmetry in particle-hole channel 
which are compatible with tetragonal symmetry  such 
as $  V  \cos 2\phi \cos 2\phi^\prime$,  $W  \sin 2\phi \sin 2\phi^\prime$. 
If such interactions were present the $B_{1g}$ and /or $B_{2g}$ channels surely would be influenced by vertex corrections.

Regarding the discrepancies on the shape of $B_{1g}$  peak structure between the theory and the experiment, 
we presume  that  there exists a very strong inelastic scattering acting in $B_{1g}$ channel  which suppresses and broadens the 
$B_{1g}$ peak. 
The introduction of inelastic scattering acting in $B_{1g}$ channel  would not modify the low frequency behaviors
since they are  mostly dictated by the symmetry of superconducting order parameter.

Borrowing the arguments in Sec. IV A, we may expect that the effect of inelastic scattering in  $B_{1g}$ channel  would be strongest near
$\omega=2 (\Delta_g+\Delta_s)$. Then, the weak peak structure  at $\omega=2 (\Delta_g+\Delta_s)$  of  \textit{ renormalized }
$A_{1g}$   
spectra  [See Fig. 4]  would be suppressed, and a peak would show up again at $\omega=2(\Delta_g-\Delta_s)$  in agreement with
the experiment.

The physical origin of the hypothetical  inelastic scattering acting in $B_{1g}$ channel  is not clear, but the Raman experiment
[ Fig. 5 (a) ]    strongly suggests  that   the scattering between $B_{1g}$ phonons and electrons might be a possible candidate.

We also note that the inclusion of the detailed momentum anisotropy of SG compatible with the tetragonal symmetry
would give the results which  will agree better with the experiments.

In summary, we have studied the electronic Raman scattering of Borocarbide superconductors based on the weak coupling theory
with $s+g$-wave gap symmetry.   The low frequency behaviors and the relative peak positions can be naturally understood within
a simple one-loop calculation. However, there are discrepancies in the detailed shape of peak structures between theoretical predictions and 
experimental results. The inclusion of  the strong inelastic scattering between  $B_{1g}$ phonon and electron may resolve the 
discrepancies.

\begin{acknowledgments}
We are grateful to In-Sang Yang for useful comments and allowing us to reproduce their data in our paper.
Fruitful discussions with Hee Sang  Kim and Tae Suk  Kim are also acknowledged.
This work was  supported by the Korea Science and Engineering
Foundation (KOSEF) through the grant No. 1999-2-11400-005-5, and by the 
Ministry of Education through Brain Korea 21 SNU-SKKU Program.
\end{acknowledgments}
\appendix
\section{\label{omega} Low frequency  behaviors of correlation functions}
At low frequency, the angular integral of Eq. (\ref{novertex}) is dominated by the contributions from the angular  region
where $\Delta(\phi) \sim 0$. In the region, the gap function $\Delta(\phi)$ can be linearized with respect to the  angle $\phi_0$ defined by
$ \Delta(\phi_0)=0$.

Let us first consider a  pure $d$-wave case  $\Delta(\phi)=\Delta_d  \cos 2\phi$.
If we take into account the contributions from the first quadrant only, the gap function vanishes at $\phi=\pi/4$.
Writing $\phi=\pi/4+x$, $|x| \ll 1$, the integral of Eq. (\ref{novertex}) can be reduced to
\begin{equation}
-\textrm{Im} \pi^R_{\gamma \gamma} \sim \frac{N_F}{\omega}\,
\int_{-\frac{\omega}{4 \Delta_d}}^{\frac{\omega}{4 \Delta_d}}\,\frac{d x}{2\pi}
\frac{ \gamma^2(\phi=\frac{\pi}{4}+x) [4 \Delta_d^2 x^2]}{
\sqrt{\omega^2-16 \Delta_d^2 x^2}}.
\end{equation}
The vertex factors become [see Eq. (\ref{vertexfactor})]
\begin{eqnarray}
\gamma_{A_{1g}}(\frac{\pi}{4}+x) &\sim & -1, \\
\gamma_{B_{1g}}(\frac{\pi}{4}+x) &\sim & -2 x, \\
\gamma_{B_{2g}}(\frac{\pi}{4}+x) &\sim & 1.
\end{eqnarray}
Using  scaling $x=\frac{\omega}{4 \Delta_d} y$, we obtain for $\omega \ll 2 \Delta_d$
\begin{eqnarray}
\label{dwavecase}
-\textrm{Im} \pi^{R (D)}_{A_{1g}}(\omega) & \sim & \omega,  \nonumber\\
-\textrm{Im} \pi^{R (D)}_{B_{1g}}(\omega) & \sim & \omega^3,  \nonumber\\
-\textrm{Im} \pi^{R (D)}_{B_{2g}}(\omega) & \sim & \omega,
\end{eqnarray}
where the superscript $(D)$ indicates the symmetry of superconducting order parameter.
The results Eq. (\ref{dwavecase}) coincide with those obtained by Devereaux and Einzel.\cite{devereaux1}

Next let us consider the $s+g$-wave case $\Delta(\phi)=\Delta_s+\Delta_g \cos 4\phi$. 
Upon linerization, one can write
\begin{equation}
\Delta(\phi=\phi_0+x) \sim - a x, \quad
a=4 \Delta_g  \sin 4 \phi_0,
\end{equation} 
where the angle $\phi_0$ is defined by the relation
\begin{equation}
\Delta_s+\Delta_g \cos 4\phi_0=0,\quad \Delta_s,\Delta_g > 0.
\end{equation}
Then, the angular integral can be expressed as:
\begin{equation}
-\textrm{Im} \pi^R_{\gamma \gamma} \sim \frac{N_F}{\omega}\,
\int_{-\frac{\omega}{2 a}}^{\frac{\omega}{2a}}\,\frac{d x}{2\pi}
\frac{ \gamma^2(\phi=\phi_0+x) [a^2 x^2]}{
\sqrt{\omega^2-4 a^2 x^2}}.
\end{equation}
The vertex factors become
\begin{eqnarray}
\gamma^2_{A_{1g}}(\phi_0+x) &\sim & \cos^2 4 \phi_0=(\Delta_s /\Delta_g)^2 \\
\label{sgb1vertex}
\gamma^2_{B_{1g}}(\phi_0+x) &\sim & \frac{1-\Delta_s/\Delta_g}{2}-[2 \sin 4\phi_0] x \nonumber \\
&-&[4 \cos 4 \phi_0 ] x^2 \\
\gamma^2_{B_{2g}}(\phi_0+x) &\sim & \frac{1+\Delta_s/\Delta_g}{2}.
\end{eqnarray}
In Eq. (\ref{sgb1vertex}) the second term is odd in $x$,  thus it gives the vanishing result upon integration.
The first and the third term  of Eq.  (\ref{sgb1vertex}) dominate the low energy behavior. 
Carrying out the integrals by rescaling as in $d$-wave case we get
\begin{eqnarray}
\label{lowAcase}
-\textrm{Im} \pi^{R (SG)}_{A_{1g}}(\omega) & \sim & \gamma^2_{A_1}\,(\Delta_s/\Delta_g)^2 \, \omega,  \\
\label{lowB1case}
-\textrm{Im} \pi^{R (SG)}_{B_{1g}}(\omega) & \sim &  \gamma^2_{B_1}\,[\frac{1-\Delta_s/\Delta_g}{2}] \,
\omega  \nonumber \\
&+& \textrm{const.} \omega^3,  \\
\label{lowB2case}
-\textrm{Im} \pi^{R (SG)}_{B_{2g}}(\omega) & \sim &  \gamma^2_{B_2}\,[\frac{1+\Delta_s/\Delta_g}{2}] \,\omega,
\end{eqnarray}
where the superscript $(SG)$ indicates the symmetry of order parameter. 
The coefficients of vertex factors $\gamma^2_{A_1},\, \gamma^2_{B_1}, \,\gamma^2_{B_2}$  are reinstated for clarity.
The above results differ from the pure $d$-wave case by the $\omega$-linear component of $B_{1g}$ spectra.  
We also note the dependence on the gap ratio $0 \le \Delta_s/\Delta_g \le 1 $. 
For the small gap ratio, the $\omega$-linear component of $A_{1g}$ spectra  is suppressed, while that of 
$B_{1g}$ spectra is enhanced. This agrees well with the numerical result. [See Fig. \ref{novertexresult} and Fig. \ref{experimentfit}]
\section{Calculation of  vertex correction}
First, write
\begin{equation}
\label{vertexdefinition}
\Gamma_{2,3}(i\omega,\vp)=\Gamma_{2,3}^s(i\omega)+\Gamma_{2,3}^g(i\omega) \cos 4\phi.
\end{equation}
Define two-component column vectors,
\begin{equation}
\Gamma_2(i\omega)=\left [
\begin{array}{c}
 \Gamma_2^s(i\omega) \cr 
 \Gamma_2^g(i\omega) 
\end{array} \right ]
,\quad
\Gamma_3(i\omega)=\left [ 
\begin{array}{c}
 \Gamma_3^s(i\omega) \cr 
 \Gamma_3^g(i\omega) 
\end{array}  \right ].
\end{equation}
Then, the integral equation for the vertex correction Eq. (\ref{vertexcorrection}) can be recast in a matrix form.
\begin{equation}
\label{matrixequation}
\left [
\begin{array}{cc} I_2 + A & -B \cr 
-B & I_2-C 
\end{array} \right ]
\left[
\begin{array}{c} \Gamma_2 \cr \Gamma_3  \end{array} \right ]=
\left [\begin{array}{c} 0 \cr \tilde{\gamma}  \end{array}  \right ],
\end{equation}
where $I_2$ is the 2x2 identity matrix, and $A,B,C$ are  2x2 matrices to be defined below.
$\tilde{\gamma}$ is a two-component column vector defined by
\begin{eqnarray}
\tilde{\gamma}&=& \left [ \begin{array}{c} 1 \cr 0  \end{array} \right ], \;\; \textrm{for} \;\; \gamma=1,   \nonumber \\
\tilde{\gamma}&=& \left [ \begin{array}{c}  0 \cr 1 \end{array} \right ], \;\;\textrm{for} \;\; \gamma=\cos 4\phi.
\end{eqnarray}
The matrix equation  Eq.  (\ref{matrixequation})  can be solved, yielding the desired solution:
\begin{eqnarray}
\label{vertexsolution}
\Gamma_{2 \gamma}&=&\Big[  (I_2-C) B^{-1} (I_2+A)-B\Big ]^{-1} \tilde{\gamma}, \nonumber \\
\Gamma_{3 \gamma}&=&\Big[   (I_2-C)-B(I_2+A)^{-1} B \Big ]^{-1} \tilde{\gamma}.
\end{eqnarray}
The 2x2  matrices $A,B,C$ are defined as follows:
\begin{equation}
A=\left [ \begin{array}{cc}  a_1 & a_2 \cr a_3 & a_4  \end{array}  \right ],\;\;
B=\left [ \begin{array}{cc}  b_1 & b_2 \cr b_3 & b_4  \end{array}  \right ],\;\;
C=\left [ \begin{array}{cc}  c_1 & c_2 \cr c_3 & c_4  \end{array}  \right ].
\end{equation}
The matrix elements are given by [$p=(i\epsilon,\vp)$]
\begin{eqnarray}
a_1(i\omega)&=&\frac{T}{N} \sum_{i\epsilon,\vp}\,I_s(\phi) A(p,i\omega), \nonumber \\
a_2(i\omega)&=&\frac{T}{N} \sum_{i\epsilon,\vp}\,I_s(\phi) (\cos 4 \phi ) A(p,i\omega), \nonumber \\
a_3(i\omega)&=&\frac{T}{N} \sum_{i\epsilon,\vp}\,I_g(\phi) A(p,i\omega), \nonumber \\
a_4(i\omega)&=&\frac{T}{N} \sum_{i\epsilon,\vp}\,I_g(\phi) (\cos 4 \phi ) A(p,i\omega), \nonumber \\
b_1(i\omega)&=&\frac{T}{N} \sum_{i\epsilon,\vp}\,I_s(\phi) B(p,i\omega), \nonumber \\
b_2(i\omega)&=&\frac{T}{N} \sum_{i\epsilon,\vp}\,I_s(\phi) (\cos 4 \phi ) B(p,i\omega), \nonumber \\
b_3(i\omega)&=&\frac{T}{N} \sum_{i\epsilon,\vp}\,I_g(\phi) B(p,i\omega),\nonumber \\
b_4(i\omega)&=&\frac{T}{N} \sum_{i\epsilon,\vp}\,I_g(\phi) (\cos 4 \phi ) B(p,i\omega), \nonumber \\
c_1(i\omega)&=&\frac{T}{N} \sum_{i\epsilon,\vp}\,I_s(\phi) C(p,i\omega), \nonumber \\
c_2(i\omega)&=&\frac{T}{N} \sum_{i\epsilon,\vp}\,I_s(\phi) (\cos 4 \phi ) C(p,i\omega), \nonumber \\
c_3(i\omega)&=&\frac{T}{N} \sum_{i\epsilon,\vp}\,I_g(\phi) C(p,i\omega), \nonumber \\
c_4(i\omega)&=&\frac{T}{N} \sum_{i\epsilon,\vp}\,I_g(\phi) (\cos 4 \phi ) C(p,i\omega),
\end{eqnarray}
where 
\begin{eqnarray}
\label{couplings}
& &I_s(\phi)=\lambda_s+\lambda_{sg} \cos 4\phi,\;
I_g(\phi)=\lambda_g \cos 4 \phi+ \lambda_{sg}, \nonumber \\
& &\lambda_s=-N_F D_s,\; \lambda_g=-N_F D_g,\;\lambda_{sg}=-N_F D_{sg}.
\end{eqnarray}
The vertex correction Eq. (\ref{vertexsolution}) together with the Eq. (\ref{vertexdefinition}) should be substituted into 
the integral Eq. (\ref{susceptibility}).
Note that the solutions  Eq. (\ref{vertexsolution})  depend only on the external frequency $i\omega$, so that just an analytic 
continuation  $i\omega \to \omega+i \delta$   without additional integraions   is sufficient to obtain the final answer.
\section{Choice of Coupling constants}
The gap equation Eq. (\ref{gapeq}) can be written as a set of coupled equations. Notations are defined in Eq. (\ref{couplings}).
\begin{eqnarray}
\label{thegapeq}
\Delta_s&=&\Delta_s ( \lambda_s I_0 + \lambda_{sg} I_1) + \Delta_g ( \lambda_s + \lambda_{sg} I_2), \nonumber \\
\Delta_g&=&\Delta_s ( \lambda_g I_1 + \lambda_{sg} I_0) + \Delta_g ( \lambda_g I_2 + \lambda_{sg} I_1),  \nonumber \\
I_n&=&\int_0^{\omega_c} d \xi_{\vp} \int_0^{\pi} \frac{d x}{\pi} (\cos x)^n\,
\frac{E_{\vp}/2T}{E_\vp}.
\end{eqnarray}
Solving Eq. (\ref{thegapeq}) at $T=T_c$, we obtain a relation between $T_c$ and coupling constants.
\begin{equation}
\label{eqtc}
\ln \big[ 1.135  \frac{\omega_c}{T_c} \big]=
\frac{(\frac{\lambda_g}{2}+\lambda_s)-\big[
(\frac{\lambda_g}{2}-\lambda_s)^2+2 \lambda^2_{sg} \big]^{1/2}}{\lambda_s \lambda_g-\lambda_{sg}^2}.
\end{equation}
For a pure $g$-wave case, the well-known result $T_c=1.135\times \omega_c e^{-2/\lambda_g}$ is reproduced.
Fixing the cutoff $\omega_c=100 \textrm{mev}$, we have chosen the following two sets of the coupling constants
by solving Eq. (\ref{thegapeq}, \ref{eqtc}).
\begin{eqnarray}
& &(\textrm{I}) \lambda_s=0.13,\;\;\lambda_g=0.22,\;\; \lambda_{sg}=0.15, \nonumber \\
& &\Delta_s=1.34 \textrm{mev},\;\; \Delta_g=1.72 \textrm{mev},\;T_c=15.9 \textrm{K}.
\end{eqnarray}
\begin{eqnarray}
& &(\textrm{II}) \lambda_s=0.05,\;\;\lambda_g=0.42,\;\; \lambda_{sg}=0.06, \nonumber \\
& &\Delta_s=0.43 \textrm{mev},\;\; \Delta_g=2.5 \textrm{mev},\;T_c=14.2 \textrm{K}.
\end{eqnarray}

\end{document}